\documentclass[aps,pra,floatfix,amsmath,amssymb,eqsecnum,nofootinbib,twocolumn,superscriptaddress]{revtex4-2}

\usepackage{subcaption}
\usepackage{graphicx}
\usepackage{hyperref}
\usepackage{algpseudocodex}
\usepackage{listings}
\usepackage{natbib}
\usepackage{tikz}
\usepackage{amsthm}
\usepackage{ragged2e}
\usetikzlibrary{quantikz2, positioning, calc}

\definecolor{classiq_red_fill}{HTML}{F5CED7}
\definecolor{classiq_red_border}{HTML}{F43764}
\definecolor{classiq_blue_fill}{HTML}{98EBEF}
\definecolor{classiq_blue_border}{HTML}{119DA4}
\definecolor{classiq_green_fill}{HTML}{E5F79C}
\definecolor{classiq_green_border}{HTML}{92AC2D}

\theoremstyle{definition}
\newtheorem*{theorem}{Theorem}
\newtheorem*{definition}{Definition}

\newcommand{\qctrl}{\textnormal{ctrl}}

\begin{document}

\title{Efficient Quantum Control via Automatic Control Skips}

\author{Peleg Emanuel}
\affiliation{Classiq Technologies}

\author{Eyal Cornfeld}
\affiliation{Classiq Technologies}

\author{Ravid Alon}
\affiliation{Classiq Technologies}

\author{Shmuel Ur}
\affiliation{Classiq Technologies}

\author{Israel Reichental}
\affiliation{Classiq Technologies}

\date{\today}

\begin{abstract}
    Control of quantum operations is a crucial yet expensive construct for quantum computation.
    Efficient implementations of controlled operations often avoid applying control to certain subcircuits, which can significantly reduce the number of gates and overall circuit depth.
    However, these methods are specialized and circuits frequently need to be implemented manually.
    This paper presents a generic method for finding "skippable" patterns without having to tailor implementations for each algorithm. We prove that finding the optimal operations to be skipped is generally NP-hard. Nevertheless, sub-optimal, polynomial approximation algorithms that find skippable subcircuits can lead to over $50\%$ improvement in circuit metrics for real-world applications.
\end{abstract}

\maketitle

\section{Introduction}

Conditional statements are an integral part of imperative programming languages, where they are needed to implement almost any classical algorithm. Similarly, their quantum counterpart, known as quantum control, is key to implementing quantum algorithms~\cite{nielsen2010quantum}. 
Classically, conditional statements are implemented by executing different code blocks according to the Boolean value of some expression previously evaluated. In quantum computing, evaluating a Boolean condition typically requires measuring a qubit, which collapses its wavefunction. Since this destroys the coherence needed for quantum computation, such measurements are highly discouraged. Instead, one may use unitary controlled operations, which preserve coherence.

Formally, controlled operations are defined as follows. Let $U$ be a unitary operation to be conditionally applied to the target wavefunction $|\psi_{t} \rangle \in \mathcal{H}_{target}$. Let $|\psi_{c} \rangle = \alpha |0\rangle + \beta |1\rangle \in \mathcal{H}_{control}$ be the state of a Boolean control qubit, where the Hilbert spaces $\mathcal{H}_{control}, \mathcal{H}_{target}$ are disjoint. $\qctrl \left(1, U\right)$ acts on the combined Hilbert space $\mathcal{H}_{control} \oplus \mathcal{H}_{target}$, such that $U$ is applied only if $\psi_c$ is in the $|1\rangle$ state, namely,
\begin{equation}
    \qctrl \left(1, U\right) |\psi_{c}\rangle |\psi_{t}\rangle
    = \alpha |0\rangle |\psi_{t} \rangle + \beta |1\rangle |U \psi_{t} \rangle
    .
\end{equation}
The generalization to $\qctrl\left(s, U\right)$, with general control states of arbitrary length and value, is straightforward.

Control gates preserve coherence and, with it, the key advantages of quantum computation. However, implementing controlled gates incurs significant overhead. To begin with, a single-qubit control acting on a single-qubit gate will turn it into a two-qubit gate. The application of two-qubit gates is more erroneous and time-consuming in almost any quantum computer architecture~\cite{doi:10.1126/science.abb2823, Krantz_2019, trapped_ions, neutral_atoms}. Furthermore, multi-controlled gates are not natively implemented by quantum hardware and must be decomposed into sequences of single- and two-qubit gates. Fault-tolerant quantum computation requires additional decomposition into the \emph{Clifford+T} gate set. T-gates are particularly costly for error correction, although recent methods suggest their impact can be substantially mitigated~\cite{magicstatecultivation}. 
The T-gate count, T-depth, and Clifford two-qubit gate count are all significantly affected when controlled operations are applied~\cite{maslov_mcx, Selinger:2013ksm}.
Given the shortage of quantum computational resources, the efficiency of quantum programs is of great concern. 
Hence, the search for better implementations of controlled and multi-controlled operations is an active field of research~\cite{Huang_MCX, Rosa_MCX, Vale_MCX}.

It is often possible to skip the control of some operations, which greatly improves circuit quality. Most notably, one may skip controlling operations that form a conjugation. For example, 
$\qctrl\left(U^\dagger V U\right) = U^\dagger \qctrl\left(V\right) U$
for any unitaries $U, V$. The control state is omitted for brevity,
and we refer to $U, U^\dagger$ as a \emph{conjugation pair}. Such patterns are ubiquitous in quantum algorithms due to the reversible nature of unitary computation. One may often need to perform intermediate calculations ($U$) before the desired computation ($V$), then \emph{uncompute} 
$U$ to release resources ($U^\dagger$)~\cite{uncomputation}. For example, ripple carry addition can be implemented as a sequence of nested conjugations~\cite{ripple-carry}, and the control of the entire circuit is then reduced to that of a single CX gate.

\begin{figure}[!b]
    \begin{quantikz}[row sep=0.4cm]
        & & \ctrl{2} & & & & \ctrl{2} & & &\\
        & & & & \ctrl{1} & & & & \ctrl{1} &\\
        & \gate{A} & \targ{} & \gate{B} & \targ{} & \gate{C} & \targ{} & \gate{D} & \targ{} & \gate{E}
    \end{quantikz}
    \caption{
    \justifying
    Gray code pattern. Control of the CX gates may be skipped.}
    \label{fig:gray-code}
\end{figure}
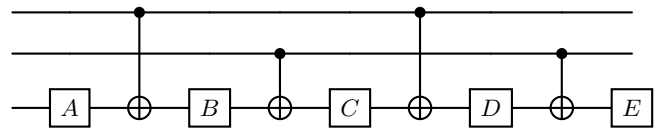

Conjugation pairs are not the only pattern that can be skipped in quantum control. Let $U$ be a unitary that can be decomposed as an ordered product of $n$ unitary operations, $U = \prod_{1 \le j\le n} \mathcal{U}_j$, with multiplication performed in the ascending order of the indices.
Let $S \subseteq \left[1, n\right]$ be a subset of these operations, such that $\prod _{j \in S} \mathcal{U}_j = 1$. It is straightforward to verify that control of the operations at $j \in S$ may be skipped~\cite{equation2verify}, namely,
\begin{equation}
    \qctrl \left(U\right) = \prod _{j}
    \begin{cases}
        \mathcal{U}_j & j \in S, \\
        \qctrl \left( \mathcal{U}_j\right) & \textnormal{else}. 
    \end{cases}
    \label{eq: skip S}
\end{equation}
Conjugations are the special case of $|S| = 2$. The above property allows the construction of specialized control for numerous quantum algorithms. For example, a standard subroutine that often appears in algorithms based on Gray codes~\cite{nielsen2010quantum, Selinger:2013ksm, maslov_mcx} is depicted in Fig.~\ref{fig:gray-code}. Control of all four CX gates may be skipped.

Identifying \emph{skippable} sets of operations can significantly improve the resulting quantum circuits. 
However, the search space is immense, since there are exponentially many operation subsets. Nevertheless, high-level descriptions of quantum programs often allow for a short decomposition of circuits. Furthermore, potential subsets can frequently be identified from the functional intent of the program, as in the case of conjugation pairs originating from computations and their uncomputations~\cite{Spire}. See, for example, the specialized control of modular addition, considered by Beauregaurd in Ref.~[\onlinecite{shor1}].

A low-level description of the circuit could expose additional skippable subsets, and thus introduce the potential for further improvements hidden in higher-level descriptions. Suppose, for example, an X gate appears twice on the same qubit. The two operations form a conjugation pair since $\textnormal{X}^2 = 1$, and can be skipped during circuit control regardless of the role of each operation in the program. This pattern may appear only for a specific implementation of a subroutine and is hence independent of the algorithmic intent; it would not be identified in a high-level description. Gate-level decompositions generally contain considerably more operations, enlarging the search space and making it hard to find candidate subsets. Moreover, scalable quantum programming requires a high-level description~\cite{goldfriend2025designsynthesisscalablequantum}, which makes manual optimization at the gate level impractical.

In this paper, we report on our recent work on automatically identifying possible skippable subsets~\cite{control_patent}. We prove that finding optimal subsets is generally NP-hard and present approximate algorithms that significantly improve real-world circuits.
Essentially, we describe a method to generically construct specialized implementations.
The method applies to any hierarchical description, yet works best with a low-level description. It can be used in addition to manual high-level optimizations and fits within the framework of a quantum program compiler.

\section{Problem formulation}
Let $U = \prod_{1 \le j\le n} \mathcal{U}_j$ be a decomposition of $U$ into unitaries; we want to find an efficient implementation for $\qctrl\left(U\right)$.
As mentioned above, the number of possible subsets is exponential in $n$. Given a subset $S$, verification that $\prod_{j \in S} \mathcal{U}_j = 1$ generally requires the computation of the unitary product. 
Therefore, we restrict ourselves to a subset of the possible solutions, where verification is easy: subsets comprised of conjugation pairs. We will soon find that the special case is NP-hard, nonetheless.

\begin{figure*}[!t]
    \begin{quantikz}[row sep=0.4cm]
        & \ctrl{1} & \ctrl{1} & \ctrl{1} & \ctrl{1} \\
        & \targ{} & \ctrl{1} & \targ{} & \ctrl{1} \\
        & & \targ{} & & \targ{}
    \end{quantikz}
    =
    \begin{quantikz}[row sep=0.3cm]
        & & \ctrl{1} & & \ctrl{1} \\
        & \targ{} & \ctrl{1} & \targ{} & \ctrl{1} \\
        & & \targ{} & & \targ{}
    \end{quantikz}
    =
    \begin{quantikz}[row sep=0.4cm]
        & \ctrl{1} & & \ctrl{1} & \\
        & \targ{} & \ctrl{1} & \targ{} & \ctrl{1} \\
        & & \targ{} & & \targ{}
    \end{quantikz}
    \caption{
        \justifying
        Circuits implementing the same controlled gate $ \qctrl\left(1, U\right)$,
        with $U = \mathcal{U}_a \mathcal{U}_b \mathcal{U}_a ^\dagger  \mathcal{U}_b ^\dagger$, $\mathcal{U}_a = \text{X} \otimes \text{I}$
        and $\mathcal{U}_b = \text{CX}$.
        From left to right, the naively controlled circuit (all gates are controlled), skipped $\mathcal{U}_a$ and $\mathcal{U}_a^\dagger$, and
        skipped $\mathcal{U}_b$ and $\mathcal{U}_b^\dagger$.
}
\label{fig:equivalent-control}
\end{figure*}
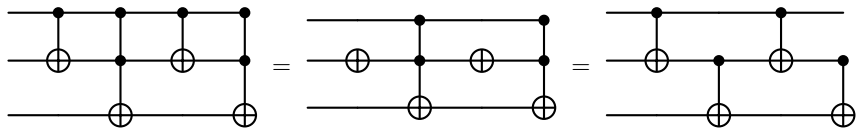

Denote by $\mathfrak{P}$ the set of all possible $\mathcal{O}\left(n^2\right)$ conjugation pairs in the decomposition of $U$. Identifying $\mathcal{U}_j ^\dagger$ is immediate both at the gate level and in a high-level description, where inversion is often integral to the language and written explicitly~\cite{qmod}. Therefore, it is straightforward to identify $\mathfrak{P}$. Now, we look for subsets of $\mathfrak{P}$ that multiply to 1. To start, consider two conjugation pairs,
$\mathcal{U}_a, \mathcal{U}_a ^\dagger$ and $\mathcal{U}_b, \mathcal{U}_b ^\dagger$ that form the set $S=\{\mathcal{U}_a, \mathcal{U}_a ^\dagger,\mathcal{U}_b, \mathcal{U}_b ^\dagger\}$. If one pair, say $a$, precedes the other in the circuit, $\prod_{j \in S} \mathcal{U}_j = \mathcal{U}_a \mathcal{U}_a ^\dagger \mathcal{U}_b \mathcal{U}_b ^\dagger = 1$. However, if the pairs "interlace," namely, appear in the order $\mathcal{U}_a, \mathcal{U}_b, \mathcal{U}_a^\dagger, \mathcal{U}_b^\dagger$, the multiplication
$\prod_{j \in S} \mathcal{U}_j = \mathcal{U}_a \mathcal{U}_b \mathcal{U}_a ^\dagger  \mathcal{U}_b ^\dagger = 1$ {iff} $\left[\mathcal{U}_a, \mathcal{U}_b\right] = 0$. 
This observation naturally generalizes to larger subsets of $\mathfrak{P}$. Pairs $a, b \in \mathfrak{P}$ can belong to the same skippable subset $\mathfrak{P}_{*}$ only if the operations either commute or do not interlace~\cite{generalization}, allowing simple verification that a subset $\mathfrak{P}_{*}$ is indeed skippable.

Not all control skips are equally beneficial. Consider, for example,
the case of interlacing pairs with $\mathcal{U}_a = \text{X} \otimes \text{I}$ and $\mathcal{U}_b = \text{CX}$. 
Either may be skipped, but not both, since $\left[\mathcal{U}_a, \mathcal{U}_b\right] \neq 0$.
Possible implementations are depicted in Fig.~\ref{fig:equivalent-control}.
If one wishes to minimize the number of two-qubit gates, it is better to skip $\mathcal{U}_b$ and avoid introducing CCX gates, which decompose into at least three CX gates, even if one utilizes relative phases~\cite{maslov_mcx}. Therefore, we consider a weighted optimization problem, which we now formally define.

\begin{definition}
    Let $\mathfrak{P} = \{{\left(i, j, w\right)} \mid {i < j}, {\mathcal{U}_j = \mathcal{U}_i ^\dagger}, {w \in \mathbb{R}}\}$ be the weighted set of conjugation pairs from the unitary decomposition $U = \prod_{1\le j \le n} \mathcal{U}_j$. 
    \textbf{Max Conjugation Pairs} (MCP) is the problem of finding a subset $\mathfrak{P}_{*} \subset \mathfrak{P}$ maximizing $\sum_{a \in \mathfrak{P}_{*}} w\left(a\right)$ such that for any $a, b \in \mathfrak{P}_{*}$, either $\left[\mathcal{U}_{i\left(a\right)}, \mathcal{U}_{i\left(b\right)}\right] = 0$, or the sections $\left[i\left(a\right), j\left(a\right)\right]$ and $\left[i\left(b\right), j\left(b\right)\right]$ do not overlap.
\end{definition}

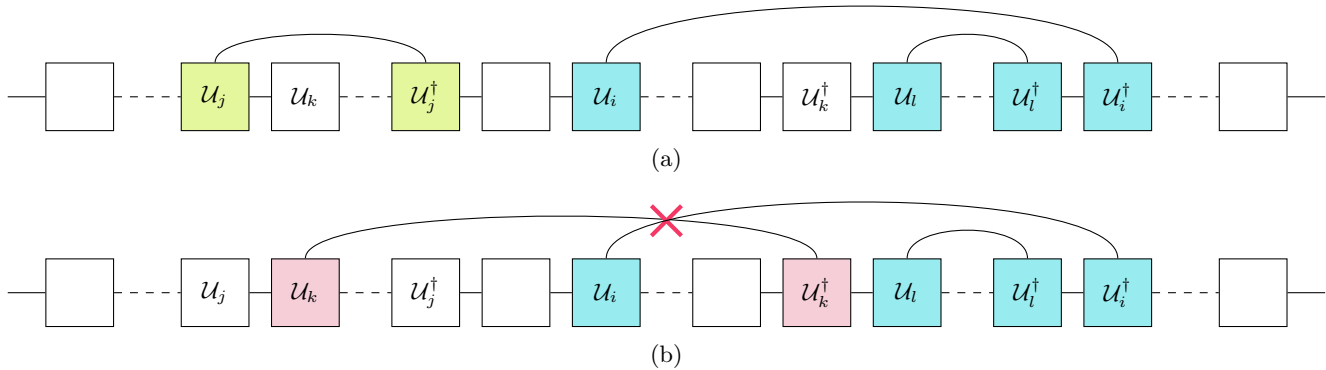
\begin{figure*}[!t]
    \centering
    \foreach \i in {1,2} {
        \begin{subfigure}{\textwidth}
            \centering
            \begin{tikzpicture}[
                box/.style={draw=black, minimum width=0.9cm, minimum height=0.9cm, fill=none}
            ]
    
                \def\boxgap{4.0cm}
                \def\sidegap{0.3cm}

                \ifnum\i=1
                    \def\fillua{1}
                    \def\fillub{0}
                    \def\drawillegalarch{0}
                    \def\drawleftarch{1}
                \else\ifnum\i=2
                    \def\fillua{0}
                    \def\fillub{1}
                    \def\drawillegalarch{1}
                    \def\drawleftarch{0}
                \fi\fi
                
                \node[box, fill={\ifnum\fillua=1 classiq_green_fill\else none\fi}] (A) at (0,0) {$\mathcal{U}_{j}$};
                \node[box] (AL) at ($(A) - (\sidegap + 1.5cm,0)$) {};
                \node[box, fill={\ifnum\fillub=1 classiq_red_fill\else none\fi}] (AR) at ($(A) + (\sidegap + 0.9cm,0)$) {$\mathcal{U}_{k}$};

                \node[box] (B) at ($(A) + (\boxgap,0)$) {};
                \node[box, fill={\ifnum\fillua=1 classiq_green_fill\else none\fi}] (BL) at ($(B) - (\sidegap + 0.9cm,0)$) {$\mathcal{U}_{j}^{\dagger}$};
                \node[box, fill={classiq_blue_fill}] (BR) at ($(B) + (\sidegap + 0.9cm,0)$) {$\mathcal{U}_{i}$};
                
                \node[box, fill={\ifnum\fillub=1 classiq_red_fill\else none\fi}] (C) at ($(B) + (\boxgap,0)$) {$\mathcal{U}_{k}^{\dagger}$};
                \node[box] (CL) at ($(C) - (\sidegap + 0.9cm,0)$) {};
                \node[box, fill={classiq_blue_fill}] (CR) at ($(C) + (\sidegap + 0.9cm,0)$) {$\mathcal{U}_{l}$};

                \node[box, fill={classiq_blue_fill}] (D) at ($(C) + (\boxgap,0)$) {$\mathcal{U}_{i}^{\dagger}$};
                \node[box, fill={classiq_blue_fill}] (DL) at ($(D) - (\sidegap + 0.9cm,0)$) {$\mathcal{U}_{l}^{\dagger}$};
                \node[box] (DR) at ($(D) + (\sidegap + 1.5cm,0)$) {};
                
                \draw ($(AL.west) + (-0.5,0)$) -- (AL.west); % incoming
                \draw[dashed] (AL.east) -- (A.west);
                \draw (A.east) -- (AR.west);
                \draw[dashed] (AR.east) -- (BL.west);
                \draw (BL.east) -- (B.west);
                \draw (B.east) -- (BR.west);
                \draw[dashed] (BR.east) -- (CL.west);
                \draw (CL.east) -- (C.west);
                \draw (C.east) -- (CR.west);
                \draw[dashed] (CR.east) -- (DL.west);
                \draw (DL.east) -- (D.west);
                \draw[dashed] (D.east) -- (DR.west);
                \draw (DR.east) -- ++(0.5,0); % outgoing

                \ifnum\drawillegalarch=1
                    \draw ($(AR.north)$) .. controls +(0,0.75) and +(0,0.75) .. ($(C.north)$);
                    \coordinate (Xpoint) at ($(CL.north) + (-0.8,0.5)$);

                    \draw[classiq_red_border, ultra thick] ($(Xpoint) + (-0.2,0.2)$) -- ($(Xpoint) + (0.2,-0.2)$);
                    \draw[classiq_red_border, ultra thick] ($(Xpoint) + (-0.2,-0.2)$) -- ($(Xpoint) + (0.2,0.2)$);
                \fi
                \ifnum\drawleftarch=1
                    \draw ($(A.north)$) .. controls +(0,0.5) and +(0,0.5) .. ($(BL.north)$);
                \fi
                \draw ($(BR.north)$) .. controls +(0,1) and +(0,1) .. ($(D.north)$);
                \draw ($(CR.north)$) .. controls +(0,0.5) and +(0,0.5) .. ($(DL.north)$);
                
            \end{tikzpicture}
            \caption{}
            \ifnum\i=1
                \label{subfig:legal-conjugation-pairs}
            \else
                \label{subfig:illegal-conjugation-pairs}
            \fi
        \end{subfigure}
    }
    \caption{
    \raggedright
    Illustration of conjugation pair choices for MCP. The pairs $\mathcal{U}_{i}$, $\mathcal{U}_{i}^{\dagger}$ and $\mathcal{U}_{l}$, $\mathcal{U}_{l}^{\dagger}$ can be chosen together to skip their control operations. In \ref{subfig:legal-conjugation-pairs}, the candidate pair $\mathcal{U}_{j}$, $\mathcal{U}_{j}^{\dagger}$ is valid, and it can be added to the list of operations to be skipped. On the other hand, as shown in \ref{subfig:illegal-conjugation-pairs}, the pair $\mathcal{U}_{k}$, $\mathcal{U}_{k}^{\dagger}$ cannot be skipped together with $\mathcal{U}_{i}$, $\mathcal{U}_{i}^{\dagger}$, $\mathcal{U}_{l}$, and $U_{l}^{\dagger}$.
    }
    \label{fig:conjugation-pair-selection}
\end{figure*}

\begin{theorem}
    MCP $\in$ NP-hard.
\begin{proof}
    We prove the theorem by reduction from the Max Weighted Independent Set (MWIS) problem~\cite{lamm2018exactlysolvingmaximumweight}.
    Let $G = \left(V, E\right)$ be a graph and $\{w_v\}_{v \in V}$ a set of associated weights. Without loss of generality, we enumerate $V = \left[n\right]$, and for simplicity, we assume $G$ is connected.
    Define a circuit on $|E|$ qubits labeled by $e \in E$, and the following operations for $i \in \left[n\right]$,
    \begin{equation*}
        g_i = \prod_{e \in E}
        \begin{cases}
            \text{X}_e & i = \min e, \\
            \text{Z}_e & i = \max e, \\
            \text{I}_e & i \notin e. \\
        \end{cases}
    \end{equation*}
    and ${\mathcal{U}_i = \exp \left(\theta_i g_i\right)}$ for ${\theta_i \in \left(0, 2\pi\right)}$. Note that ${\left[\mathcal{U}_i, \mathcal{U}_j\right] = 0}$ {iff} ${\{i, j\} \notin E}$:
    if ${\{i, j\} \notin E}$, there is no qubit on which $g_i$ and $g_j$ both act nontrivially. If 
    ${\{i, j\} \in E}$, there is exactly one such qubit $e \equiv \{i, j\}$, and ${\left[X_e, Z_e\right] \neq 0}$.
    
    Consider $U = \big(\prod_i \mathcal{U}_i\big) \cdot \big(\prod_i \mathcal{U}_i^\dagger\big)$,
    with the weight $w_i$ assigned to each pair $\mathcal{U}_i, \mathcal{U}_i ^\dagger$.
    There are exactly $n$ conjugation pairs and any two pairs interlace. Hence, a solution of MCP on $U$ is a subset of commuting operators. As shown above, all operations in $\{\mathcal{U}_i\}_{i \in S}$ commute {iff} $S \subseteq V$ is an independent set, concluding the proof.
\end{proof}
\end{theorem}

In our definition of MCP, there is no restriction on the decomposition of $U$, which makes it suitable for all hierarchical description levels. However, at the gate level, one usually decomposes the circuit using a finite set of (possibly parametric) basis gates, imposing constraints on $\mathcal{U}_i$-s. We note that the proof holds if we impose such a restriction.
The parametric basis gates in the above proof are given by
\begin{equation*}
    h_{kl} \left(\theta\right) = \exp \left(i \theta \underbrace{Z \otimes Z \otimes \dots}_{k \textnormal{ times}} \otimes \underbrace{X \otimes X \otimes \dots}_{l \textnormal{ times}}\right),
\end{equation*}
with $1 \le k + l \le d_{\textnormal{max}}$ where $d_{\textnormal{max}}$ is the maximal degree in the graph, allowing for $\mathcal{O}\left(n\right)$ basis gates in the general case.
However, MWIS is NP-hard even for graphs of maximal degree 3~\cite{berman1995approximation}, for which the resulting $h_{kl}$ satisfy $k +l \le 3$. Thus, the same proof holds even if the number of basis gates is limited to nine and basis gates act on up to three qubits.

Despite the problem being NP-hard, there are still many ways to find skippable subsets that could yield significant improvements in circuit quality.
For example, MCP is polynomially reducible to MWIS. Given an MCP problem, we can define a weighted graph by creating a vertex for each conjugation pair and assigning it the respective weight of the pair, adding edges between vertices if their corresponding conjugation pairs cannot be simultaneously chosen for a skippable set.
MWIS has long been studied and has numerous approximation algorithms and successful methods for solutions on real-world graphs~\cite{berman1995approximation, SAKAI2003313, lamm2018exactlysolvingmaximumweight, xiao_efficient}; these could also be applied to MCP. This is one of many possible approaches to the problem. In the next section, we consider another algorithm and present its results.

\section{Results}
The complexity of MCP is reduced significantly if no operations in the sequence commute~\cite{no_commute}. In this case, two conjugation pairs cannot be chosen {iff} they interlace, and the problem may be understood in purely geometric terms. 

\begin{figure*}[t!]
    \begin{subfigure}{1.0\textwidth}
        \centering
        \includegraphics[width=1.0\linewidth]{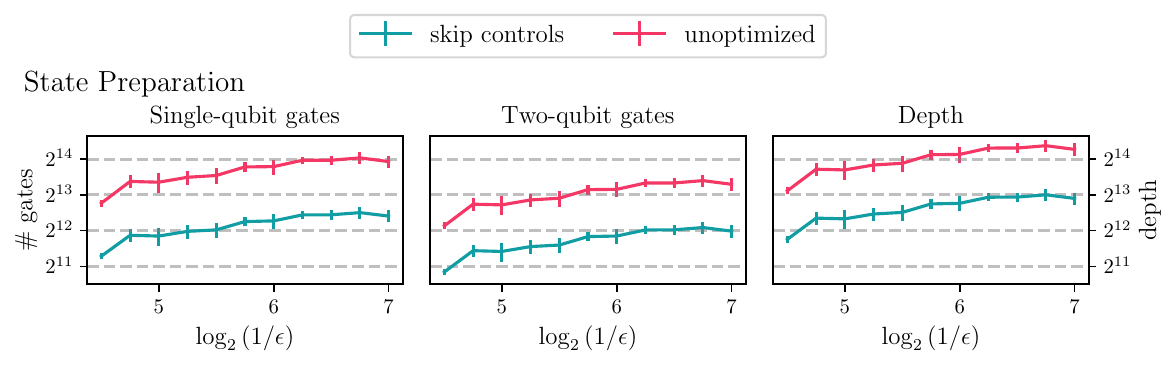}
    \end{subfigure}
    \begin{subfigure}{1.0\textwidth}
        \centering
        \includegraphics[width=1.0\linewidth]{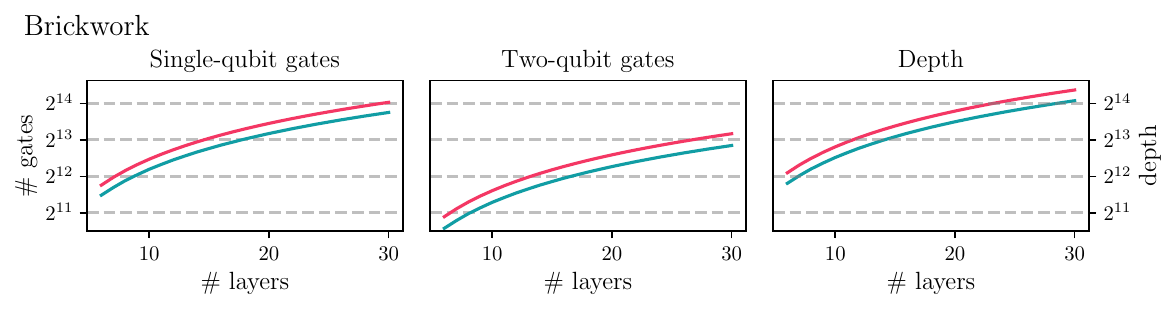}
    \end{subfigure}
    \caption{
    \justifying Circuit metric comparison of naively controlled circuits, and control-skip controlled circuits, using the non-commuting dynamic programming algorithm solving Eq.~\eqref{eq: algo-max}.
    Top, state preparation on $5$ qubits vs. error upper bound $\epsilon$, with $5$ instances for each value of $\epsilon$. Probabilities are taken from a uniform distribution on $2^5$ values and then normalized. The circuit is constructed using the built-in state preparation of \texttt{Qmod}~\cite{qmod}, based on the algorithm in Ref.~[\onlinecite{grover2002creatingsuperpositionscorrespondefficiently}]. Bottom, brickwork random circuit~\cite{random_quantum_circuits}, for different numbers of layers. Each "brick" is a two-qubit random unitary.
    Transpilation is done with \texttt{Qiskit} 1.2.4~\cite{javadiabhari2024quantumcomputingqiskit} and the basis gates
    $\{
        \textnormal{X}, \textnormal{Y}, \textnormal{Z},
        \textnormal{T}, \textnormal{T}^\dagger, \textnormal{S}, \textnormal{S}^\dagger,
        \textnormal{SX}, \textnormal{SX}^\dagger, \textnormal{H}, \textnormal{CX},
        \textnormal{P}
    \}$.
    }
    \label{fig: results}
\end{figure*}

We place the operations in the decomposition of $U$ on a line in chronological order and draw an arc connecting operations of any chosen conjugation pair, going above the operator sequence, as illustrated in Fig.~\ref{fig:conjugation-pair-selection}. Suppose a pair with operations at indices $i$ and $j$ is chosen. Other arcs must lie within the range $\left[i + 1, j-1\right]$ or outside it, but cannot cross the $ij$ arc. This allows us to divide the problem into smaller subproblems and construct a recursion relation for the maximum weight attainable in the range $\left[i, j\right]$, $W_{i, j}$.

Let us traverse the circuit from the left. Suppose there is an operator at index $j > 1$ such that $\mathcal{U}_j = \mathcal{U}_1^\dagger$. We may choose the pair $\mathcal{U}_1, \mathcal{U}_j$ for the optimal sequence. Then, other conjugation pairs must belong to the solution of MCP on either $\left[2, j-1\right]$ or $\left[j + 1, n\right]$. Alternatively, suppose no conjugation pair with $\mathcal{U}_1$ is chosen, then the solution of MCP on $\left[1, n\right]$ is equal to that on $\left[2, n\right]$. Proceeding with the above procedure for $\mathcal{U}_2$, $\mathcal{U}_3$ etc, we obtain the recursion relation,
\setcounter{equation}{2}
\begin{widetext}
\begin{equation}
    \label{eq: algo-max}
        W_{i, j} = \max \bigg \{W_{i + 1, j},
        w \left(\mathcal{U}_i\right)
        + \max _{k \in \left[i+1, j\right]. \mathcal{U}_k = \mathcal{U}_i ^{-1}}
        \left(W_{i+1, k} + W_{k +1, j}\right) \bigg \},
\end{equation}
\end{widetext}
with $w$ the weight function, satisfying $w \left(\mathcal{U}\right) = w \left(\mathcal{U}^\dagger\right)$.
The solution to the non-commuting MCP problem is given by $W_{1,n}$
and is straightforwardly attainable via dynamic programming~\cite{cormen01introduction}. The calculation of the maximum is $\mathcal{O}\left(r\right)$, with $r$ the maximal multiplicity of an operator in the sequence. There are $\mathcal{O}\left(n^2\right)$ subranges of $\left[1, n\right]$, so the algorithm's runtime complexity is $\mathcal{O}\left(rn^2\right)$. Generally, $r = \mathcal{O}\left(n\right)$, resulting in a total $\mathcal{O}\left(n^3\right)$. However, $r$ is typically significantly lower than $n$, and may also be artificially bounded, making runtime complexity quadratic at the price of an approximate solution.

The premise of non-commuting operators is implausible in real-world circuits since operations acting on different qubits necessarily commute. Nevertheless, the above algorithm still outputs a valid solution to the general problem. We consider its effectiveness. The results of the algorithm are displayed in Fig.~\ref{fig: results}
for two classes of random circuits: state preparation with random probabilities, for different error upper bounds $\epsilon$, and random brickwork circuits~\cite{random_quantum_circuits}, for different numbers of layers. The data was obtained using the following procedure. First, one generates a circuit, as described separately for each case in the figure caption. Then, the circuits are transpiled. A solution of $W_{1, n}$ is obtained using Eq.~\eqref{eq: algo-max}, with the weights $w\left(\textnormal{CX}\right)=3$ and $w\left(g \neq \textnormal{CX}\right) = 1$.
The resulting controlled circuit is retranspiled.
Due to the last step, efficient control reduces the number of single-qubit gates, in addition to that of two-qubit gates.

For random brickwork circuits, skipping controls reduces gate counts and depth by $\sim 8 \%$, while for state preparation, reduction may be over $50 \%$.
That being said, the algorithm presented is not the optimal way to control either state preparation algorithms or brickwork circuits. Knowledge of the intricate details of each algorithm can allow us to tailor better implementations. Rather, a significant improvement may be obtained even \emph{without} such knowledge.
As quantum programs increase in size and complexity, generic optimization methods prove more useful and are key to implementation at scale.

\section{Extensions}
For simplicity, circuits are described at the gate level for most examples in this paper. Nevertheless, we stress that the MCP problem and its approximating algorithms are defined for an arbitrary decomposition and hierarchy level. As part of a compiler architecture, MCP algorithms can be applied to high-level, low-level, and any intermediate circuit representation, possibly achieving different levels of improvement.

The choice of decomposition can greatly affect the success of MCP algorithms, even within the same hierarchical level. Consider, as an example, a qubit on which the sequence
$\textnormal{X} \dots \textnormal{Z} \textnormal{H}$ appears. There appear to be no conjugation pairs, yet, recalling $\textnormal{X} = \textnormal{H} \textnormal{Z} \textnormal{H}$ and $\textnormal{H}^2 = 1$, we can rewrite the sequence as $\textnormal{X} \dots \textnormal{H} \textnormal{X}$, revealing a possible conjugation pair. Thus, it is best to decompose the circuit with either $\textnormal{X}$ or $\textnormal{Z}$ in the basis gate set, but not both. Similarly, many parametric gates $g\left(\theta\right)$ form an additive group, such that $g \left(\theta_1\right) g\left(\theta_2\right) = g\left(\theta_1 + \theta_2\right)$. It may be useful to decompose 
$g \left(\theta_1\right) \dots g \left(\theta_2\right) 
= g \left(\theta_1\right) \dots g \left(\theta_1\right) ^\dagger g \left(\theta_1 + \theta_2\right)$, exposing another possible conjugation pair.
We leave the exploration of optimizations over unitary decompositions that yield better MCP solutions for future work.

\section{Conclusions}
Efficiently performing controlled operations is crucial for the effective execution of quantum programs. We present a new approach for generically and automatically constructing specialized control implementations.

Our approach fits perfectly within the framework of a quantum program compiler. It resembles
optimizations done by classical compilers in that it is blind to functional intent,
and requires no effort from the programmer. As mentioned above, automatic control skipping algorithms will probably not outperform specialized implementations but could be used on top of them. For example, one can manually choose conjugation pairs to be skipped and impose additional constraints on the conjugation pair set of the MCP problem, but not necessarily exhaust it. MCP solutions can also be searched for at lower hierarchy levels than those for which specialized implementations are defined.

We presented a single approximation for the MCP problem to show how generic control-skipping algorithms can significantly improve circuit metrics.
Introducing additional MCP approximations and decomposition methods that enhance the effectiveness of these algorithms could allow for significant improvements in controlled operation quality, which is essential for the scalable implementation of quantum algorithms.

\begin{acknowledgments}
    The authors thank Ron Hass, Yehuda Naveh, Nathaniel A. Rosenbloom, and Ori Roth for insightful discussions and notes on earlier versions of this manuscript.
\end{acknowledgments}

\bibliography{references, notes}

\begin{thebibliography}{29}
\bibitem [{equ()}]{equation2verify}%
  \BibitemOpen
  \href@noop {} {}\bibinfo {note} {We note that both sides of Eq.~\ref{eq: skip S} are identical regardless of the control state. If the control state is 1, the left-hand side reads $U$, and the right-hand side reads $\prod_j \mathcal{U}_j$, equal by the definition of $U$. If the control state is $0$, the left-hand side reads $1$ and the right-hand side $\prod_{j \in S} \mathcal{U}_j$, equal by the choice of $S$.}\BibitemShut {Stop}%
\bibitem [{gen()}]{generalization}%
  \BibitemOpen
  \href@noop {} {}\bibinfo {note} {This may be proved by induction over the number of interlacing pairs $n$. Let $S$ be a set of conjugation pairs such that if two pairs interlace, they commute. First, iteratively cancel all pairs $\mathcal{U}_j, \mathcal{U}_j ^\dagger$ such that no operation in $S$ appears between $\mathcal{U}_j$ and $\mathcal{U}_j ^\dagger$. For $n=0$ this yields $\prod_{j \in S} \mathcal{U}_j = 1$, proving the induction base. For $n>1$, there must be two pairs $\mathcal{U}_a$ and $\mathcal{U}_b$ such that the pattern $\mathcal{U}_a \mathcal{U}_b \dots \mathcal{U}_a ^\dagger$ appears, and $\mathcal{U}_b ^\dagger$ is appears somewhere either before $\mathcal{U}_a$ or after $\mathcal{U}_a ^\dagger$. Since these pairs interlace, $\mathcal{U}_a$ and $\mathcal{U}_b$ commute, and we may switch them without altering $\prod_{j \in S} \mathcal{U}_j$. Doing so, the pairs of $\mathcal{U}_a$ and $\mathcal{U}_b$ no longer interlace, we're left with a set of $n-1$ interlacing pairs, which multiplies to
  $1$ by induction.}\BibitemShut {Stop}%
\end{thebibliography}

\end{document}